\documentclass[aps,prl,showpacs,preprintnumbers,twocolumn, amsmath, amssymb, notitlepage,superscriptaddress]{revtex4-1}

\usepackage{graphicx}
\usepackage{dcolumn}
\usepackage{bm}
\usepackage{array} 

\usepackage{pdfsync} 
\usepackage{multirow}

\usepackage{enumitem}

\def\app#1#2{%
  \mathrel{%
    \setbox0=\hbox{$#1\sim$}%
    \setbox2=\hbox{%
      \rlap{\hbox{$#1<$}}%
      \lower1.1\ht0\box0%
    }%
    \raise0.25\ht2\box2%
  }%
}

\def\app#1#2{%
  \mathrel{%
    \setbox0=\hbox{$#1\sim$}%
    \setbox2=\hbox{%
      \rlap{\hbox{$#1>$}}%
      \lower1.1\ht0\box0%
    }%
    \raise0.25\ht2\box2%
  }%
}

\newcommand{\ket}[1]{\left| #1 \right>} 

\setlist{nolistsep} 

\begin{document}

\title{Tunable Landau-Zener transitions in a spin-orbit-coupled Bose-Einstein condensate}
\author{Abraham J. Olson}
\email{olsonaj@purdue.edu}
\affiliation{Department of Physics, Purdue University, West Lafayette IN 47907}%
\author{Su-Ju Wang}
\affiliation{Department of Physics, Purdue University, West Lafayette IN 47907}%
\author{Robert J. Niffenegger}
\affiliation{Department of Physics, Purdue University, West Lafayette IN 47907}%
\author{Chuan-Hsun Li}
\affiliation{School of Electrical and Computer Engineering, Purdue University, West Lafayette IN 47907}
\author{Chris H. Greene}
\affiliation{Department of Physics, Purdue University, West Lafayette IN 47907}%
\author{Yong P. Chen}%
\email{yongchen@purdue.edu}
\affiliation{Department of Physics, Purdue University, West Lafayette IN 47907}%
\affiliation{School of Electrical and Computer Engineering, Purdue University, West Lafayette IN 47907}
\affiliation{Birck Nanotechnology Center, Purdue University, West Lafayette IN 47907}

\date{\today}
						
\begin{abstract}
The Landau-Zener (LZ) transition is one of the most fundamental phenomena in quantum dynamics. It describes nonadiabatic transitions between quantum states near an avoided crossing that can occur in diverse physical systems.  Here we report experimental measurements and tuning of LZ transitions between the dressed eigenlevels of a synthetically spin-orbit (SO) coupled Bose-Einstein condensate (BEC). We measure the transition probability as the BEC is accelerated through the SO avoided crossing, and study its dependence on the coupling between the diabatic (bare) states, eigenlevel slope, and eigenstate velocity---the three parameters of the LZ model that are independently controlled in our experiments. Furthermore, we performed time-resolved measurements to demonstrate the breaking-down of the spin-momentum locking of the spin-orbit coupled BEC in the nonadiabatic regime, and determine the diabatic switching time of the LZ transitions. Our observations show quantitative agreement with the LZ model and numerical simulations of the quantum dynamics in the quasimomentum space.  The tunable LZ transition may be exploited to enable a spin-dependent atomtronic transistor. \end{abstract}

\maketitle
\section{Introduction}
Controllable ``synthetic'' gauge fields can be created using laser-dressed adiabatic states in ultracold atomic gases \cite{Dalibard_RMP_2011,Galitski_Nat_2013}. Rapid experimental progress has, among many other developments, realized measurements of both bosonic and fermionic ultracold atoms in synthetic spin-orbit (SO) gauge fields \cite{Lin_Nature_2011, Cheuk_PRL_2012,Wang_PRL_2012}. Such developments have motivated many recent proposals for using more elaborate laser-dressed synthetic gauge fields to create quantum simulators using ultracold atoms \cite{Bloch_NatPhys_2012} to realize novel quantum states such as topological insulators \cite{Goldman_PRL_2010, Beri_PRL_2011} and Majorana fermions~\cite{Zhang_PRL_2008,Zhu_PRL_2011}. 

For the laser-dressed synthetic gauge fields realized in experiments and proposed in theories, it is typically assumed that the system adiabatically follows the dressed eigenlevels \cite{Galitski_Nat_2013,Dalibard_RMP_2011}. Naturally, the hitherto unexplored regimes in which this adiabatic assumption no longer holds are also of interest, as more complex studies and coupling schemes proceed. A unique feature of SO gauge fields is the spin-momentum locking in which a change of momentum can yield a change in spin if the system evolves slowly enough to adiabatically follow the eigenlevel. For a SO-coupled Bose-Einstein condensate (BEC), diabatic transitions between eigenlevels correspond to a breakdown of the spin-momentum locking. This work investigates such transitions as a synthetically SO-coupled BEC is accelerated through the SO eigenlevel avoided crossing. We find that the Landau-Zener (LZ) theory provides an excellent quantitative model for understanding such transitions.

\section{Theory of LZ transitions in a SO-coupled BEC}
The Landau-Zener model \cite{Zener_PRSA_1932} describes the transition of a quantum state between two adiabatic eigenlevels when some parameter that controls the eigenstate of the system is linearly varied in time \footnote{The linear assumption often provides a good approximation near avoided crossing even for more general systems with non-linearly varying energies.}. The LZ model assumes the time-dependent Schr\"{o}dinger equation 
\begin{equation}
i \hbar \frac{\partial}{\partial t} \left( \begin{array}{c} \phi_1 \\ \phi_2 \end{array} \right) = \left( \begin{array}{cc} E_1(x) & \Omega/2 \\ \Omega/2 & E_2(x) \end{array}\right) \left( \begin{array}{c} \phi_1 \\ \phi_2 \end{array} \right),
\end{equation}
where the diabatic (``bare'') states, $\phi_{1,2}$, have energies, $E_{1,2}$ which linearly vary with some adiabatic parameter, $x$, and cross at $x_c$ ($E_1(x_c) = E_2(x_c)$). The difference in slopes between the energy levels at the crossing is defined as $\beta = \left|\partial E_1(x)/ \partial x - \partial E_2(x)/ \partial x \right|_{x=x_c} $. $\Omega$ is the coupling between the two diabatic energy levels. This coupling ``dresses'' the bare energy levels and forms new adiabatic eigenlevels separated by $\Omega$ at the avoided crossing.

\begin{figure*}[!t]
  \includegraphics[width=1\textwidth]{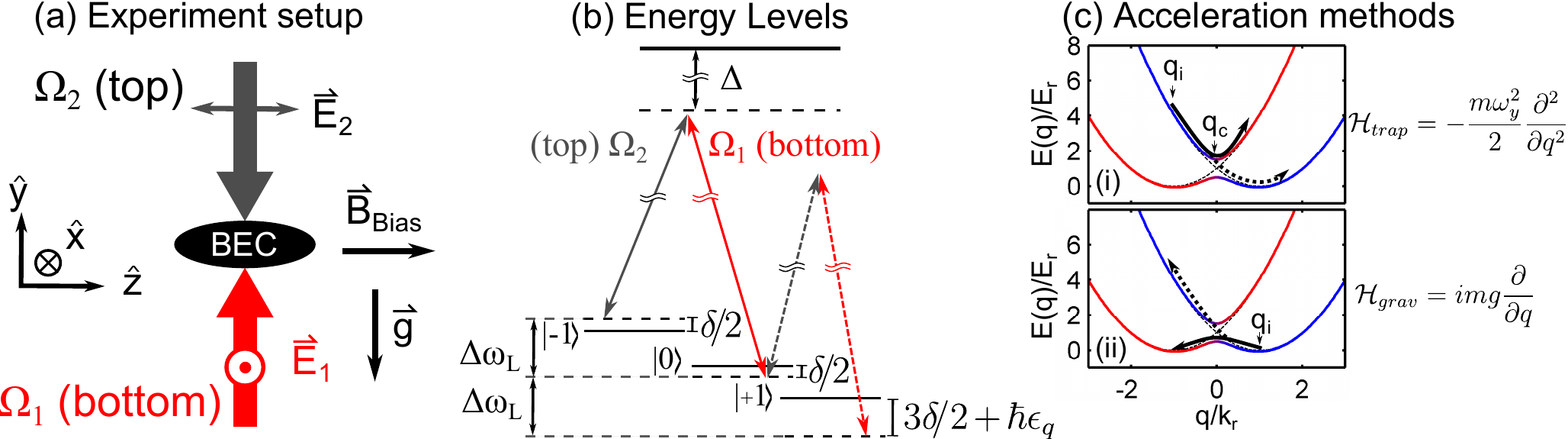}
  \caption{(Color online) (a,b) Counter-propagating, linearly polarized laser beams couple the $m_F=-1,0$ states of the $^{87}$Rb BEC via a Raman transition. An external bias magnetic field along the $\hat{z}$-direction Zeeman splits the $m_F$ states, and is used to control the Raman detuning $\delta$. Gravity acts in the $-\hat{y}$-direction. (c) Two acceleration methods used to study Landau-Zener transitions at this avoided crossing of the SO eigenlevels: (i) the acceleration induced by the force of the trapping potential drives transitions from the upper to lower dressed eigenlevel, (ii) the acceleration induced by the gravitational force drives transitions from the lower to upper dressed eigenlevel. The dashed black curves indicate two ``bare'' spin state energy levels, the  solid lines indicate the two SO-coupled eigenlevels [with the color indicating the bare state spin component, red (blue) for spin $m_F=-1$ (0)], here shown for when $\delta = 0$ $E_r$. Solid and dotted arrows depict adiabatic (intraband) motion and diabatic (interband) LZ transition, respectively. }
	\label{fig:expSetup}
\end{figure*}

If a quantum state begins in one adiabatic eigenlevel far from the avoided crossing and is given some eigenstate velocity, $v=dx/dt$, of the adiabatic parameter in the direction toward the avoided crossing, it acquires some probability, $P_{LZ}$, to make a diabatic transition to the other adiabatic eigenlevel as it moves past $x_c$. This diabatic transition probability is determined to be
\begin{equation}\label{eqn:LZprob}
P_{LZ}  = \exp \left[-2 \pi (\Omega/2)^2 / (\hbar v \beta)\right].
\end{equation}
With small velocities or strong coupling, the adiabatic theorem holds and negligible transfer occurs. However, with high velocities or weak coupling, the diabatic transition probability can be significant. 

Previous experimental measurements of LZ transitions have been performed with diverse physical systems. Some examples include ultracold atoms in accelerated optical lattices \cite{Wilkinson_Nature_1997, Zenesini_PRL_2009, Kling_PRL_2010}, Feshbach associated ultracold molecules \cite{Kohler_RMP_2006}, Rydberg atoms \cite{Rubbmark_PRA_1981}, as well as in solid-state qubits \cite{Sillanpaa_PRL_2006,Cao_NatComm_2013} and spin-transistors \cite{Betthausen_Science_2012}. In this paper, we measure the LZ transition probability of a BEC with synthetic 1D SO coupling of equal Rashba and Dresselhaus types \cite{Lin_Nature_2011}. The SO coupling is the result of adiabatic ``dressed'' states formed by Raman coupling the ``bare'' quadratic dispersion curves of two $m_F$ spin states \cite{Higbie_PRL_2002,Spielman_PRA_2009}. The coupling of the spin states by the Raman field leads to a SO coupling of the form:
\begin{equation}\label{eqn:HSO}
{\cal H}_{SO} = \left( \begin{array}{cc}
\frac{\hbar^2}{2 m} (q+k_r)^2 - \delta/2 & \Omega_R/2 \\
\Omega_R/2 & \frac{\hbar^2}{2 m} (q-k_r)^2 + \delta/2 \end{array} \right)
\end{equation}
where $\Omega_R$ is the Raman coupling strength, $\hbar k_r$ is the single-photon recoil momentum from the coupling lasers, $m$ is atomic mass, $\delta$ is the Raman detuning, $\hbar$ is the reduced Planck's constant, and $\hbar q$ is the quasimomentum ($q$ is the canonical momentum conjugate to the position coordinate $\hat y$). Applying the Landau-Zener model to a SO-coupled BEC, the adiabatic parameter is $q$, the velocity is $v=dq/dt$, the coupling strength is $\Omega_R$, and $\beta$ is defined as the magnitude of the diabatic curve slope difference, obtained by making a linear approximation of ${\cal H}_{SO}$ near the diabatic crossing point, $q_c$. 
\section{Experimental Setup}
For our experiment, we produce nearly-pure 3D BECs of $1-2 \times 10^4$ $^{87}$Rb atoms in an optical dipole trap \cite{Olson_PRA_2013}, with trapping frequencies tuned in the range of $\omega_{z,y}/ 2 \pi \approx 180$-$450$ Hz and $\omega_{x} / 2 \pi \approx 50$-$90$ Hz. To create synthetic spin-orbit coupling, we employ counter-propagating Raman beams along the $\hat{y}$-axis which couple the $\ket{m_F}$ states of the $F=1$ ground state manifold of $^{87}$Rb (see Fig.~\ref{fig:expSetup}), similar to that of Lin \emph{et al.} \cite{Lin_PRL_2009, Lin_Nature_2011}. The two Raman beams are generated from the same laser source ($\omega_L = 2 \pi {\times} 383\,240$ GHz), have a frequency difference of $ \Delta \omega_L = 2\pi {\times} 3.5$MHz, and have perpendicular linear polarizations when incident on the BEC [Fig.~\ref{fig:expSetup}(a)]. The detuning provided by the quadratic Zeeman shift on the $\ket{m_F=+1}$ state ($\epsilon_q \approx 2 \pi {\times} 3.4$ kHz) allows for the system to be approximated by the two-state description of Eq.~(\ref{eqn:HSO}). (We simplify the actual three-level $F=1$ ground state of $^{87}$Rb into the two-level SO-coupled system following the convention detailed in Refs.~\cite{Lin_Nature_2011} and \cite{LeBlanc_NJP_2013}.) The natural energy scale for the SO-coupled system is the recoil energy from the coupling laser fields, $E_r = \frac{\hbar^2 k_r^2}{2 m}=2\pi \times \hbar \times 3.75$ kHz, where $k_r = 2 \pi / \lambda $ and $\lambda = 782.26$ nm. For the experiments considered in this work, effects due to atom-atom interactions are negligible because the interaction energy of atoms in the BEC ($E_{int}\approx 0.1 E_r$) is much smaller than the canonical kinetic energy ($E_{kin}\gtrsim 4 E_r$). The dynamics of the BEC relevant for the experiments here can therefore be described by a 1D Schr\"odinger equation, where the Hamiltonian ${\cal H} = {\cal H}_{SO} + {\cal H}_{trap}$. Recalling the relation of position and momentum operators $\hat{y} = i \partial / \partial \hat{q}$, it is elucidating to express the trapping term in the y-axis as ${\cal H}_{trap} = -\frac{m \omega_y^2}{2} \frac{d^2}{dq^2}$, which shows how the trapping potential acts as a ``kinetic energy'' in quasimomentum space \cite{Higbie_PRA_2004, Zhang_PRA_2013stability}.  

\section{Measurements of $P_{LZ}$}

As schematically shown in Fig.~\ref{fig:expSetup}(c), measurements of the Landau-Zener transition probability, $P_{LZ}$, were performed by first preparing the BEC in either the upper or lower SO eigenlevel, with initial quasimomentum $\hbar q_i$ far from the SO diabatic crossing at $\hbar q_{c}$. The BEC was then accelerated \footnote{The acceleration in the real space can be either positive or negative.} through the diabatic crossing, either by the optical trapping force or by gravity. Depending on the accelerating force, the BEC acquired different eigenstate velocities, $d q / dt$, as it passed $q_{c}$ (detailed procedures of preparing different $d q /dt$ are presented in Appendix A). After the crossing and when the BEC was sufficiently far from the diabatic crossing such that the diabatic and adiabatic eigenstates matched to better than $97\%$, the Raman beams and dipole trap were instantly turned off to map the adiabatic dressed eigenstates to the bare spin states. A Stern-Gerlach field was then applied to separate the bare $m_F$ spin states in time-of-flight (TOF), and absorption images measured the population of each spin state to determine $P_{LZ}$ (which is $N_{0}/N_{tot}$ for the transitions in Fig.~\ref{fig:expSetup}c, with $N_0$ and $N_{-1}$ being the population in $m_F=0$ and $-1$ respectively, and $N_{tot}=N_{0}+N_{-1}$ the total population). 

\begin{figure}[!htpb]
  \includegraphics[width=3.1in]{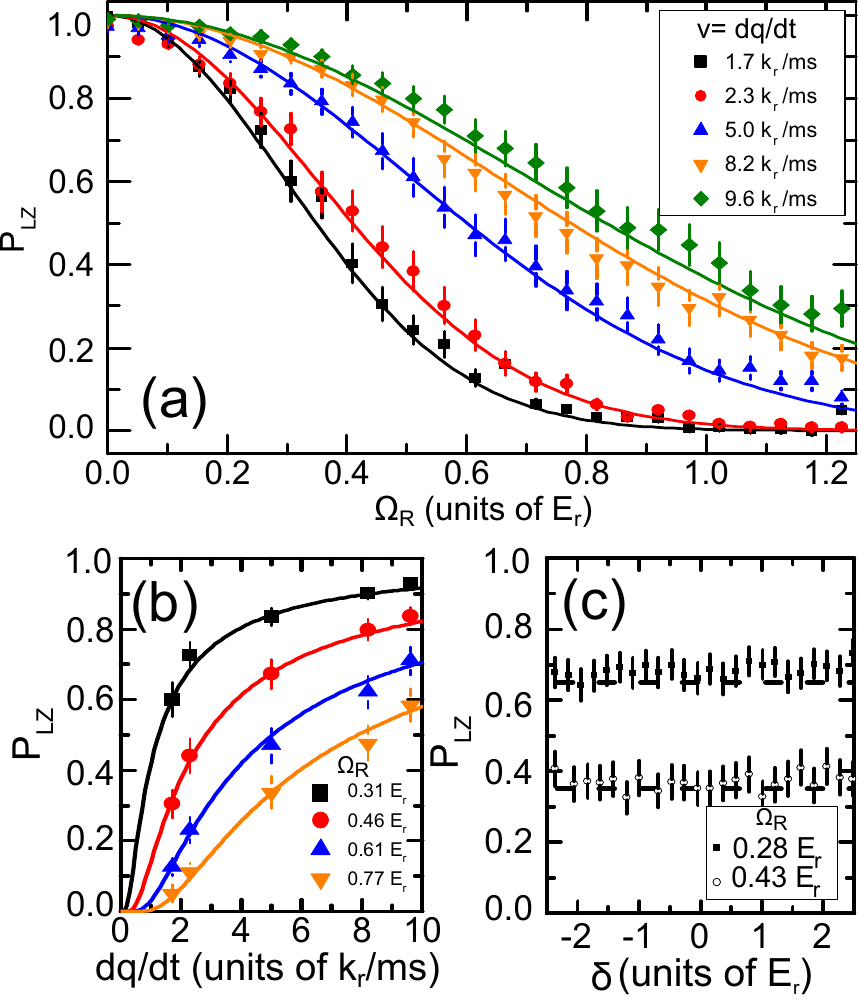}
  \caption{(Color online) (a-b) Measurement of the LZ transition probability, $P_{LZ}$, over a range of Raman coupling strengths, $\Omega_R$, and with different eigenstate velocities ($d q / d t$) at the diabatic crossing, and $\delta = 0$ $E_r$. The data in (b) is from the same experiments as (a), but plotted to show the effect of $d q / d t$ on $P_{LZ}$. All solid lines are calculated from the LZ model using Eq.~(\ref{eqn:LZprob}) and the experimental values of $\Omega_R$, $d q / d t$, and $\beta = 4 E_r/ k_r$ with no free parameters. The data with $d q / d t = 1.7$ k$_r/$ms corresponds to the case of acceleration due to gravity (Fig.~\ref{fig:expSetup}c.ii); the other $d q / d t$ data are measured by applying an optical dipole trapping force of different magnitudes (controlled by the optical trap laser power) to accelerate the BEC (Fig.~\ref{fig:expSetup}c.i). All experiments were performed with $\sim 1\times 10^4$ atoms in the BEC. Each data point is the average of 3-5 measurements, and error bars indicate an average $10\%$ uncertainty in atom number due to technical noise. (c) Measurement of the LZ transition probability over a range of Raman detuning, $\delta$, from resonance.  No discernible change of $P_{LZ}$ was observed, in agreement with the theoretical model. Measurement was performed with $d q / d t = 1.7$ k$_r/$ms (supplied by gravity) with two values of $\Omega_R$: 0.28$E_r$ (closed squares) and 0.43$E_r$ (open circles). The values calculated from Eq.~(\ref{eqn:LZprob}) are shown as dashed lines.}
	\label{fig:LZvaryVelocity}
\end{figure}

Fig.~\ref{fig:LZvaryVelocity} (a,b) shows the measurement of $P_{LZ}$ for increasing coupling strengths, $\Omega_R$, and different eigenstate velocities, $dq/dt$, with the theoretically calculated $P_{LZ}$ from Eq.~(\ref{eqn:LZprob}) shown by the solid curves. In agreement with the LZ model, the transition probability increases for smaller coupling strengths or larger  $dq/dt$. The LZ model and experimental results are in good quantitative agreement to the level of our experimental resolution of $P_{LZ}$, which is limited by technical noise in the experimental imaging.  Shown in Fig.~\ref{fig:LZvaryVelocity} (c), we also measured $P_{LZ}$ over a range of Raman detuning $\delta$. The results further validate the expected LZ behavior, where $P_{LZ}$ does not depend on $\delta$ (since $\beta$ is independent of $\delta$ for this SO-coupled system).

\begin{figure}[htpb]
  \includegraphics[width=3.3in]{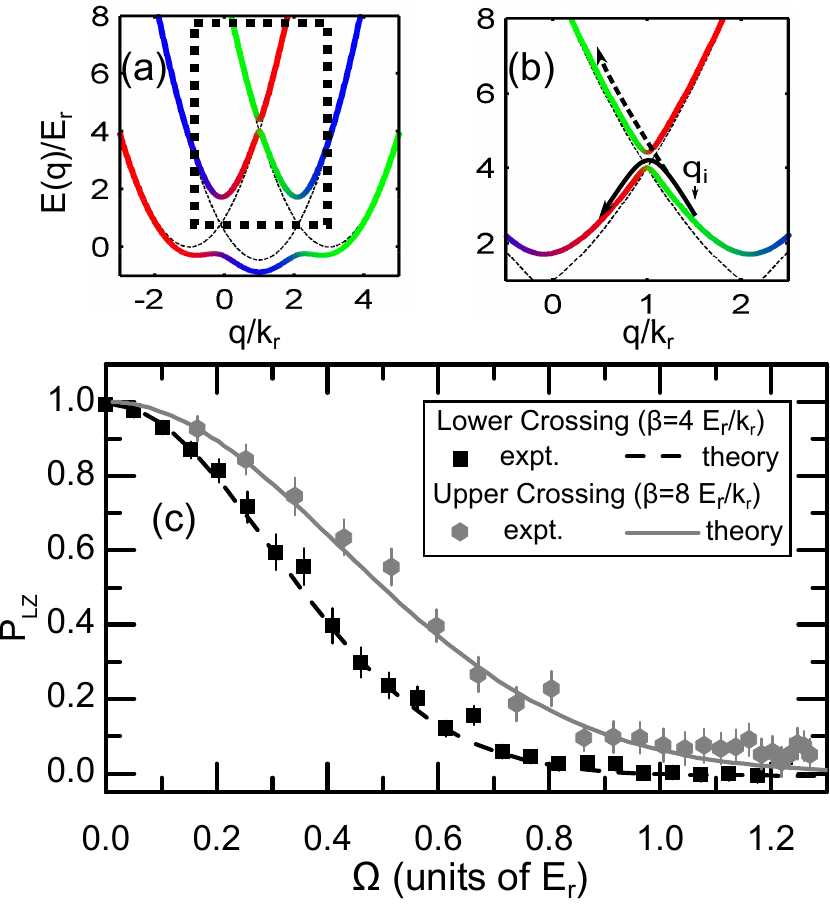}
  \caption{(Color online) Measurement of the LZ transition probability at the upper crossing of the diabatic dispersion relations for the $m_F$ spin states. (a) The full eigenlevels of the three-state system where the dashed lines indicate the bare state energy levels and the solid color lines indicate SO-coupled adiabatic eigenlevels with color representing the $m_F$ components (red for $\ket{m_F=-1}$, blue for $\ket{m_F=0}$, and green for $\ket{m_F=+1}$). The dashed box, magnified in (b), indicates the upper crossing. (c) The lower crossing data (black squares, same data shown in Fig.~\ref{fig:LZvaryVelocity}(a)) are measurements of $P_{LZ}$ for the crossing of the $\ket{m_F=-1}$ and $\ket{m_F=0}$ states where $\beta = 4$ $E_r/k_r$, and the upper crossing data (grey circles) are measurements for the crossing of the $\ket{m_F=-1}$ and $\ket{m_F=+1}$ states where  $\beta = 8$ $E_r/k_r$. The eigenstate velocity in both cases was $d q / dt = 1.7$ k$_R/$ms supplied by gravity. }
	\label{fig:LZ_changeSlope}
\end{figure}

In the SO-coupled BEC eigenlevel structure, we have the opportunity to measure $v$ and $\beta$ independently to further confirm the validity of the LZ model to this system. To measure the effect of changing $\beta$, however, it is necessary to measure $P_{LZ}$ at a different diabatic crossing. Shown in Fig.~\ref{fig:LZ_changeSlope}(a), the third-spin state in the system allows probing of the diabatic crossing of the $\ket{m_F=\pm 1}$ states where $\beta=8 E_r/k_r$, twice the value of the ground state crossing \footnote{The excited state crossing studied here is the same diabatic crossing where a ``Zitterbewegung'' was recently observed \cite{LeBlanc_NJP_2013}.}. The coupling between the $\ket{m_F=-1}$ and $\ket{m_F=+1}$ states is a four-photon process, and the strength of the coupling is numerically found to be $\Omega_{4p}/E_r \approx 0.12 (\Omega_R/E_r)^{1.75}$ \footnote{Gap $\Omega_{4p}$ is found numerically for our experimental parameters. If the $m_F=0$ state could be adiabatically eliminated, $\Omega_{4p}$ would be $\Omega_R^2 / 2(4 E_r - \hbar \epsilon_q)$, as found in \cite{LeBlanc_NJP_2013}}. To compare the two crossing in the LZ model, we present in Fig. 3(c) the $P_{LZ}$ value over a range of diabatic state coupling strengths, $\Omega$, which defines the energy gap at the avoided crossing and is $\Omega_R$ for the lower diabatic crossing and $\Omega_{4p}$ for the excited state diabatic crossing. The measurements of the excited state crossing is again in good agreement with the theoretically calculated $P_{LZ}$ from the LZ model.

\section{Time-dependent Measurements of Spin Polarization}
\begin{figure*}
\includegraphics[]{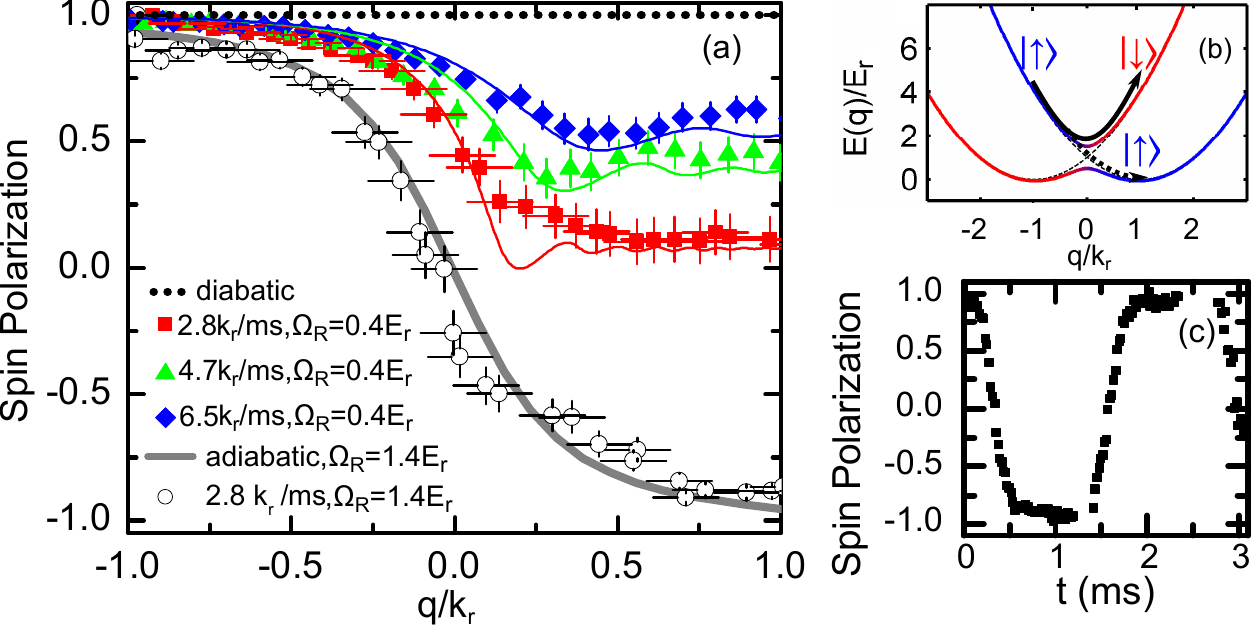}
\caption{(Color online) (a) Experimental measurements of the bare spin polarization as function of quasimomentum, q, for different degrees of adiabaticity as controlled by ($dq/dt$, $\Omega_R$), showing that less adiabatic evolutions retain less spin-momentum locking. When the BEC adiabatically follows the eigenlevel of the excited band, the spin polarization exhibits the full spin-momentum locking represented by the solid arrow in (b). Such an adiabatic evolution, shown by the open circle data in (a), was realized using $\Omega_R = 1.4 E_r$ with $dq/dt=2.8 k_r/$ms at the avoided crossing, with corresponding full spin polarization oscillations in the upper band shown in (c). Diabatic breakdown of the spin momentum locking was observed for three experimental accelerations at a lower coupling strength of $\Omega_R=0.4 E_r$ shown by the red squares, green triangles, and blue diamonds, with corresponding results from the time-dependent, 1D Schr\"{o}dinger equation simulations shown by the solid curves. Here, due to LZ transitions to the ground band the spin-momentum locking is only partially retained. For all data shown in the figure, $\delta = 0$ $E_r$, the BEC acceleration was provided by the trap as in Fig.\ref{fig:expSetup}c(i), and the error bars in (a) indicate a combination of the uncertainties in the imaging calibration and the shot-to-shot noise.}
\label{fig:LZpolVsK}
\end{figure*}

It is well known that the dressed bands (which are {\it adiabatic} energy eigenlevels) possess ``spin-momentum locking'' \cite{Zhang_PRL_2012}, where the spin composition of the dressed state is tied to its quasimomentum. Returning to a consideration of the lower SO avoided crossing, more specifically, in an adiabatically evolving SO-coupled BEC with $\delta=0$ (Fig.~\ref{fig:LZpolVsK} b), a change of the quasimomentum from $q=-$k$_r$ to $+$k$_r$ causes a flip of the bare state spins, and thus reverses the spin polarization of the BEC (where $\ket{\uparrow}\equiv \ket{m_F=0}$, $\ket{\downarrow}\equiv \ket{m_F=-1}$, spin polarization$\equiv (N_{\uparrow}- N_{\downarrow})/N_{tot}$, and $N_{tot}=(N_{\uparrow}+N_{\downarrow}$). For q sufficiently far from the avoided crossing (such that the BEC is dominantly in one bare spin state), reversing q simply reverses the spin direction. It is important to note that the spin-momentum locking, which is one of the most important general properties of spin-orbit quantum gases and underlies many novel physical effects  (such as Majorana fermions) \cite{Dalibard_RMP_2011,Zhang_PRL_2008,Zhu_PRL_2011}, is rooted in the adiabatic assumption and will break down when the adiabaticity breaks down (as in the nonadiabatic LZ transitions). 

Utilizing time, momentum and spin resolved imaging of the LZ transition process, we measure this breakdown of the SO locking. We performed such measurements by instantaneously turning off the Raman coupling during the LZ transition process (at time $t$ since the BEC starts from $q_i$ at $t=0$) and thus map the BEC dressed states on to their bare-spin component basis. These are then separated by a Stern-Gerlach pulse and imaged after TOF to determine both the bare-state spin and momentum components of the BEC \footnote{The quasi-momentum of the BEC is determined by Gaussian fitting of density profiles of both bare spin components derived from the same $q$.}. For atoms starting in the upper band with spin up, as diagrammed in Fig.~\ref{fig:LZpolVsK} (b), adiabatic evolution would lead to oscillations in the upper band of coupled spin and momentum. Nonadiabatic LZ transitions, however, cause a breakdown of the SO locking. By controlling either $\Omega_R$ or $d q / dt$ the final spin polarization of the BEC after it is accelerated across the SO avoided crossing can be controlled. 

As seen in Fig.~\ref{fig:LZpolVsK} (a), an adiabatic evolution (from a bare spin $\ket{\uparrow}$ BEC at $q=-k_r$ to a bare spin $\ket{\downarrow}$ BEC at $q=k_r$, as indicated by solid black arrow in Fig. 4 b) was realized by using an acceleration of 2.8 k$_{r}/$ms (measured at the crossing) and $\Omega_R = 1.4$E$_{r}$ where the measured (bare) spin polarization at different $q$ is shown by the open circles. The time-resolved adiabatic oscillations of the BEC's spin polarization for the same parameters is shown in Fig.~\ref{fig:LZpolVsK} (c) (with sufficiently long hold time so the BEC passes though $q=0$ three times). The breakdown of the adiabaticity, and thus full spin-momentum locking, is seen in  Fig.~\ref{fig:LZpolVsK} (a) for $\Omega_R=0.4E_r$, where different rates of BEC acceleration lead to different amounts of breakdown of the spin-momentum locking. The diabatic limit (represented by the horizontal dotted line) indicates a regime where the spin-momentum locking fully breaks down (e.g. $\Omega_R=0$E$_{r}$), and the adiabatic limit (represented by the grey solid line) reflects the spin polarization of the adiabatic band structure calculated from Eq.~\ref{eqn:HSO}. Given the range of accelerations accessible in our experimental setup, it is difficult to go from a fully adiabatic to fully diabatic regime by only tuning $dq/dt$ at a fixed $\Omega_R$. Varying $\Omega_R$, however, more easily allows for changing from adiabatic to fully diabatic. These time resolved measurements are similar to the work of Ref.~\cite{Tayebirad_PRA_2010} in optical lattices. Different from their work, however, this LZ transition is created using a Raman-coupling which gives rise to spin-dependent eigenstates, which can be ``read'' out using Stern-Gerlach separation \footnote{In addition, no real-space (e.g. optical lattice) potential is needed to realize this LZ transition.}. 

This behavior suggests using a tunable LZ transition of SO-coupled atoms to create a unique spin-dependent ``atomtronic'' device \cite{Vaishnav_PRL_2008, Pepino_PRL_2009,Beeler_Nature_2013}. This device would be an analog of a spin transistor in which $\Omega$ acts as the gate voltage, the BEC spin polarization (the output in one of the spin components) as the current, and the ``drift velocity'' $d q / dt$ induced by the force that acts as the source-drain voltage (note the qualitative similarity of Fig.~\ref{fig:LZvaryVelocity}~(b) to transistor characteristic curves). As suggested in \cite{Vaishnav_PRL_2008}, a Stern-Gerlach field then acts as a spin-filter in the readout. 

\begin{figure}[htpb]
  \includegraphics[width=3.25in]{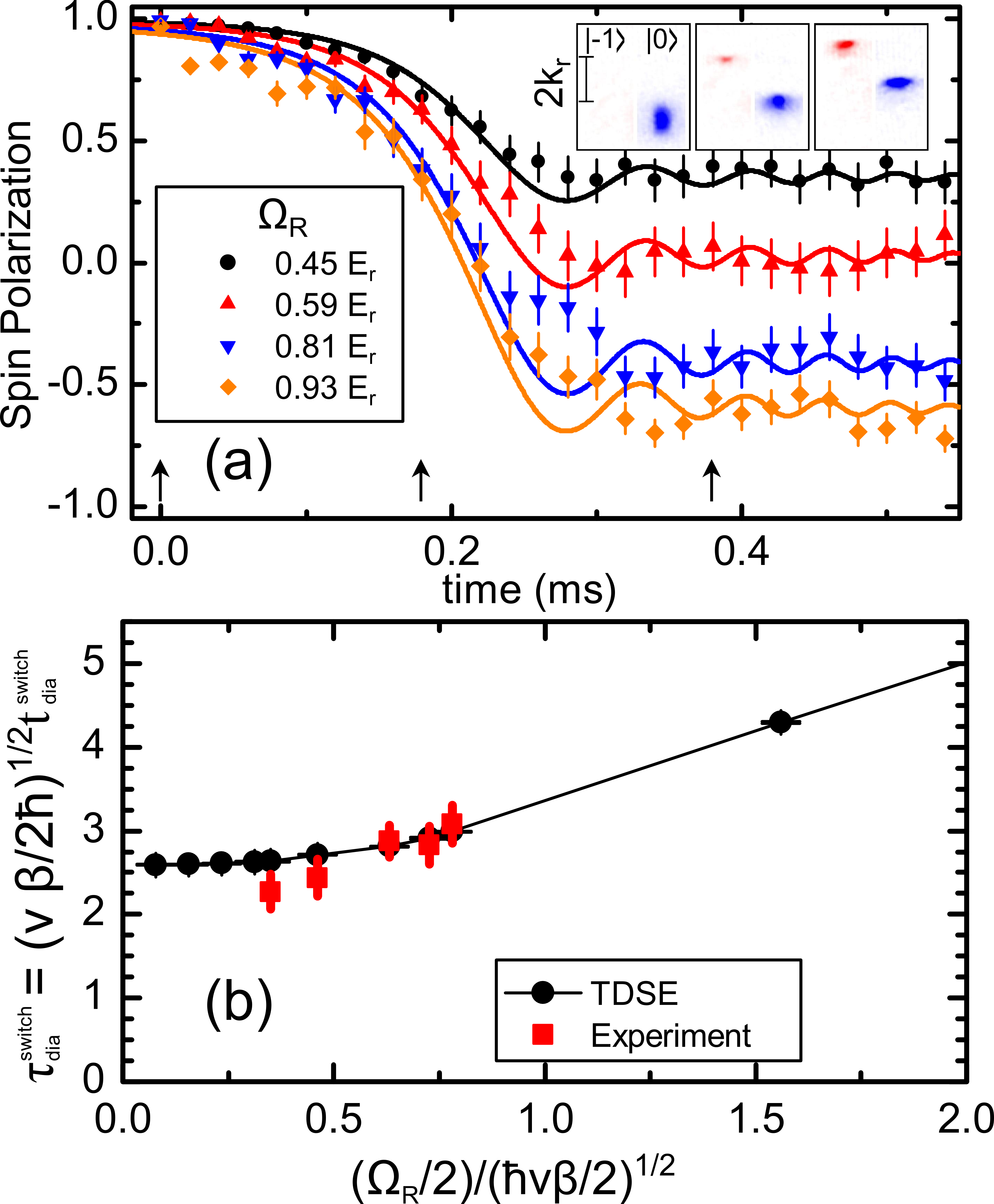}
  \caption{(Color online) (a) Measurements of the time-dependent LZ transition of the BEC passing through the diabatic crossing with $d q / dt = 5.0$ k$_r/$ms. The inset shows experimental absorption images (from left to right, at three representative points in time indicated by the arrows at $t={0,180,380}$ $\mu$s respectively for $\Omega_R = 0.59 E_r$) of the time dependent transition (false color added to distinguish the spin components). The change in BEC aspect ratio is due to a quadrupole mode excited by the state preparation process. Solid curves in (a) are from direct solution of the time dependent Schr\"odinger equation (TDSE). (b)  Measurements of the diabatic switching time, $t_{dia}^{switch}$, for different $\Omega_R$, with $d q / dt = 5.0$ k$_r/$ms, $\beta = 4$ $E_r/k_r$, and $\omega_y/ 2\pi = 338$ Hz. Here $t_{dia}^{switch}$ is found by fitting each set of data to a sigmoid function \cite{Zenesini_PRL_2009}. The switching time is scaled by $(v \beta / 2 \hbar)^{1/2}$ to compare with the theory of Ref.~\cite{Vitanov_PRA_1999}. The black dots are the results from direct solutions of the TDSE.}
	\label{fig:LZtimeDepend}
\end{figure}

An important characteristic in such devices is the ``switching time''. The switching time is the internal time of the device in which the spin is flipped, and is thus the minimum time needed to operate the spin switch. Fig.~\ref{fig:LZtimeDepend} (a) shows the resulting measurements of the BEC spin polarization as time $t$ is varied for a fixed $v$ at the crossing ($d q / dt = 5.0 k_r/$ms) and four different Raman coupling strengths. The solid lines are results of the numerical simulation of the time-dependent, 1D Schr\"{o}dinger equation (Appendix B). By fitting the experimental measurements to a sigmoid function (similar to \cite{Zenesini_PRL_2009}), we extract the time it takes to transition between the diabatic states (``switching time'', $t^{switch}_{dia}$) in Fig.~\ref{fig:LZtimeDepend}(b). We rescale the extracted switching times to $\tau^{switch}_{dia} = t^{switch}_{dia} (v \beta/2\hbar)^{1/2}$ and find agreement with Ref.~\cite{Vitanov_PRA_1999}'s predicted value of $\tau^{switch}_{dia}=2.5$ in the diabatic limit [$(\Omega_R/2)/(\hbar v \beta/2)^{1/2}<<1$]. General agreement is found with the numerical simulation in both the final spin polarization value and the timescale of the transition. Oscillatory ``quantum beats'' \cite{Garraway_PRA_1998} are seen in the numerical simulation and are the result of a coherence between the split wavepackets in the excited and lower eigenlevels after the LZ transition, but could not be conclusively observed in our data owing to our limited experimental resolution.

\section{Conclusion}
In summary, we have measured the interband LZ transition probability in a SO-coupled BEC. The coupling strength, diabatic slopes, and eigenstate velocity at the avoided crossing were each varied independently and shown to agree with the LZ prediction of Eq.~(\ref{eqn:LZprob}). We have also demonstrated the versatility of using both gravitational and optical trapping forces to prepare and drive quantum states in synthetic gauge fields. Finally, the dynamics of the transition process was directly measured and was in good agreement with our numerical calculations. The interband transitions studied in our work are entirely due to the breakdown of adiabaticity in the system, in contrast to the transitions due to 2-body collisions studied in previous experiments \cite{Williams_Science_2012, Zhang_PRA_2013stability}. As diabatic transitions are exploited in various proposals for realizing novel synthetic gauge fields in optical flux-lattices \cite{Gerbier_NJP_2010,Dalibard_RMP_2011}, Rydberg atoms \cite{Kiffner_JOPB_2013} and molecules \cite{Yarkony_RMP_1996}, our study of diabatic transitions in SO-coupled BECs may provide additional tools in designing laser-induced synthetic gauge fields. In future work, our approach could be used to probe more complex synthetic gauge fields and to observe Stueckelberg interference \cite{Kling_PRL_2010}. Another natural continuation of this work would be to study the effect of interactions on this LZ transition process \cite{Chen_NatPhys_2010}. 

\section{Acknowledgments}
The research was supported in part by DURIP-ARO Grant No. W911NF-08-1-0265, the Miller Family Endowment, the National Science Foundation (Grant PHY-1306905), and a Purdue University OVPR Research Incentive Grant. A.J.O. also acknowledges support of the U.S. National Science Foundation Graduate Research Fellowship Program. We thank Hui Zhai for helpful comments.

\appendix

\section{Appendix A: Variable eigenstate velocity}\label{sec:expVarAcceleration}

\setcounter{section}{1}
To achieve the variable eigenstate velocities at the diabatic crossing point, the BEC was first prepared in an optical dipole trap in the $\ket{m_F=0}$ state at $q = 1 k_r$. The trapping potential was instantly removed and the BEC would fall under gravity in the $\ket{m_F=0}$ diabatic state for $1.2$ ms, at which point it would reach $q \approx -1.0 k_r$. The Raman coupling was then instantly turned on along with the trapping potential, which caused the BEC to be ``accelerated back'' (decelerated in real space) through the diabatic crossing with an acceleration dependent on the trap frequency $\omega_y$. After passing the crossing, the population of the BEC in each spin component was measured to determine $P_{LZ}$. The eigenstate velocities $dq/dt$ were determined from time resolved measurements of the BEC as it crossed $q_c$, see Table~\ref{tab:variabledqdt}. [The LZ theory assumes a constant $dq/dt$, but in our system $dq/dt$ changes as the atoms are accelerated in the eigenlevels by the optical trap. We find that using the value of $dq/dt$ at the crossing result in good agreement with the theory.] For the numerical simulations of the time-dependent LZ transition shown in Fig. 4, the width ($\sigma_w$) of the initial condensate momentum distribution was set to match that of the experimentally recorded values, also shown in Table~\ref{tab:variabledqdt}.

\begingroup
\squeezetable
\begin{table}[htbp]
\begin{tabular}{*{12}{cc}}
  \hline
$\omega_y / 2 \pi$ (Hz)  &  $d q / dt$ ($k_r/$ms) & $\sigma_w$ ($k_r$) \\
\hline
264 & 2.3 & 0.31 \\
338 & 5.0 & 0.40 \\
397 & 8.2 & 0.44 \\
449 & 9.6 & 0.47\\
  \hline
\end{tabular}
\caption{The corresponding optical dipole trap frequency $\omega_y$ for each of the measured $dq/dt$ at $q=q_c$, as well as the initial momentum width of the BEC, where the momentum distribution was fitted by $p(q) =  \frac{1}{\sqrt{2 \pi} \sigma_w} \exp \left[-(q-q_i)^2/(2\sigma_w^2)\right]$.}
\label{tab:variabledqdt}
\end{table}
\endgroup

\section{Appendix B: Time-dependent Schr\"odinger equation simulation}\label{sec:tdse}

To solve the one-dimensional time-dependent Schr$\ddot{\text{o}}$dinger equation for the spin-orbit coupled BEC in a harmonic trap, we apply the Chebychev propagation method \cite{TalEzer_JChP_1984}.
\begin{widetext}
\begin{equation}
i\hbar \partial_t \Psi(q,t)=\hat{H} \Psi(q,t)=
\bigg[-\frac{1}{2}m\omega^2\frac{\partial^2}{\partial q^2}+\begin{pmatrix}
     \frac{\hbar^2}{2m}(q+k_r)^2-\delta/2& \Omega_R/2  \\
      \Omega_R/2&  \frac{\hbar^2}{2m}(q-k_r)^2+\delta/2
\end{pmatrix}\bigg]\Psi(q,t),
\end{equation}
\end{widetext}
where $\Psi(q,t)=\{\Psi_{\uparrow}(q,t),\Psi_{\downarrow}(q,t)\}^{{T}}$ is a two-component column vector written in the bare state basis $(\{|m_F=-1\rangle,|m_F=0\rangle\})$. Expanding the evolution operator in terms of the Chebychev polynomials with a renormalization of the Hamiltonian $H_{norm}$ whose eigenvalue ranges from $[\lambda_{\text{min}},\lambda_{\text{max}}]$, we arrive at
\begin{widetext}
\begin{equation}
\label{U}
\hat{U}(dt)=e^{-i\hat{H}dt/\hbar}=\sum_{n=0}^{\infty} a_n \phi_n(-i\hat{H}_{\text{norm}})=\sum_{n=0}^{\infty} a_n \phi_n\bigg(\frac{-i\hat{H}+\hat{I}(\lambda_{\text{max}}+\lambda_{\text{min}})/2}{(\lambda_{\text{max}}-\lambda_{\text{min}})/2}\bigg),
\end{equation}
\end{widetext}
where $\phi_n(x)$ is the complex Chebychev polynomial of order $n$. The expansion coefficients are
\begin{widetext}
\begin{equation}
a_n=e^{i(\lambda_{\text{max}}+\lambda_{\text{min}})dt/2\hbar} (2-\delta_{n,0}) J_n \left(\frac{(\lambda_{\text{max}}-\lambda_{\text{min}})dt}{2\hbar}\right).
\end{equation}
\end{widetext}
$J_n(x)$ is the Bessel function of order $n$. The wave function at any time is obtained by applying the evolution operator to an given initial wave function: $\Psi(q,t+dt)=\hat{U}(dt)\Psi(q,t)$. To perform the Hamiltonian operation, we represent our wave functions and operators in the Fourier discrete variable representation (Fourier-DVR) \cite{Colbert_JChP_1992}. The grid points are $q_i=q_{\text{min}}+i(q_{\text{max}}-q_{\text{min}})/(N+1)$ for $i=1,2,...,N$. We take $\{q_{\text{min}},q_{\text{max}}\}=\{-6k_r,6k_r\}$, and $N=500$. To converge the series expansion, the degree of the expansion in Eq.(\ref{U}) must be larger than $R=(\lambda_{\text{max}}-\lambda_{\text{min}})dt/2\hbar$. In our simulation, we choose the degree to be the least integer greater than or equal to $1.5R$. Since the parameter $R$ depends on $dt$, we increase the efficiency of our codes by appropriately choosing a suitable time step $(dt=0.01\hbar/E_r)$ for each time propagation.  

\begin{figure}
\includegraphics[width=3.3in]{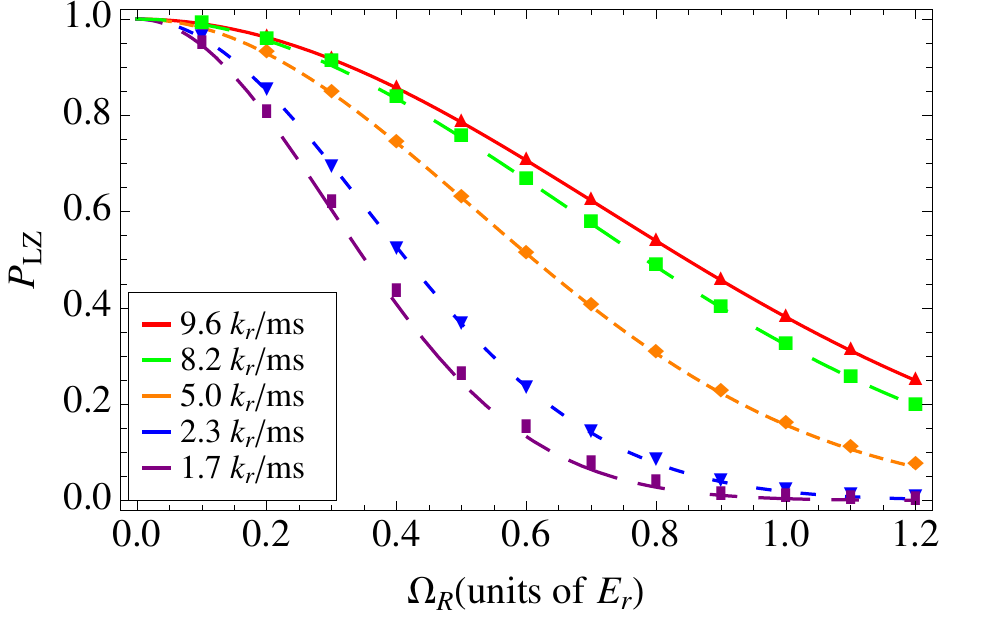}
\caption{(Color online) Comparison of the exact numerical solution (shaped symbols) and the Landau-Zener formula (solid curves) for the non-adiabatic transition probability as a function of the Raman coupling $\Omega_R$ . Different colors correspond to different eigenstate velocities at the crossing point.}
\label{fig:supplement1}
\end{figure}

In our simulation, a Gaussian wave packet in one of the adiabatic states serves as the initial wave function. Note that the adiabatic states $(\{|+\rangle, |-\rangle\})$ are related to the bare states by a unitary transformation. With the method mentioned in the first paragraph, we can evolve our system to any later time to study non-adiabatic inter-band transitions. Defining the probability for an atom to stay in $|\pm\rangle$ as $P_{\pm}(t)=\sum_{i=1}^N|\Psi_{\pm}(q_i,t)|^2$, we extract the asymptotic values of the probability for the atom to be in the other adiabatic state right after the wave packet passes the avoided crossing. 
This probability is the familiar Landau-Zener transition probability \textit{if} the energy band is linear in the bare state basis. Near the avoided crossing region, the energy bands in the spin-orbit coupled system are well described by two linear lines, and Fig.~\ref{fig:supplement1} shows that the simple LZ formula gives a very good approximation to the non-adiabatic inter-band transition probability in the spin-orbit coupled BEC. 

%

\end{document}